\documentclass[aps,prappl,reprint,a4paper,floatfix,amsmath,amssymb,amsfonts,noshowpacs,longbibliography]{revtex4-2}
\usepackage{newtxtext,newtxmath}
\usepackage[T1]{fontenc}
\usepackage[utf8]{inputenx}
\usepackage{graphicx}
\usepackage[dvipsnames]{xcolor}
\usepackage{soul}
\usepackage[textwidth=17.5cm,textheight=23.5cm,verbose,pdftex]{geometry}
\usepackage{hyperref}
\hypersetup{pdfauthor={Georg Hoffmann}, bookmarksnumbered=true, pdftitle={Suboxide desorption of elemental sources in an oxygen background: Consequences for molecular beam epitaxy and low pressure chemical vapor deposition of oxides}, colorlinks, citecolor=blue, linkcolor=blue, urlcolor=blue}

\usepackage{csquotes}

\DeclareMathAlphabet{\mathcal}{OMS}{cmsy}{m}{n}

\begin{document}
\title{Drastically enhanced cation incorporation in the epitaxy of oxides due to formation and evaporation of suboxides from elemental sources}

\author{Georg Hoffmann}
\email[Electronic mail: ]{Hoffmann@pdi-berlin.de}
\author{Zongzhe Cheng}
\author{Oliver Brandt}
\author{Oliver Bierwagen}
\email{bierwagen@pdi-berlin.de}
\affiliation{Paul-Drude-Institut f\"ur Festk\"orperelektronik, Leibniz-Institut im
Forschungsverbund Berlin e.V., Hausvogteiplatz 5-7, 10117 Berlin,
Germany}


\begin{abstract}

In the molecular beam epitaxy of oxide films, the cation (Sn, Ga) or dopant (Sn) incorporation does not follow the vapor pressure of the elemental metal sources, but is enhanced by several orders of magnitude for low source temperatures. Using line-of-sight quadrupole mass spectrometry, we identify the dominant contribution to the total flux emanating from Sn and Ga sources at these temperatures to be due to the unintentional formation and evaporation of the respective suboxides SnO and Ga$_{2}$O. We quantitatively describe this phenomenon by a rate-equation model that takes into account the O background pressure, the resulting formation of the suboxides via oxidation of the metal source, and their subsequent thermally activated evaporation. As a result, the total flux composed of the metal and the suboxide fluxes exhibit an \textsf{S}-shape temperature dependence instead of the expected linear one in an Arrhenius plot, in excellent agreement with the available experimental data. Our model reveals that the thermally activated regimes at low and high temperatures are almost exclusively due to suboxide and metal evaporation, respectively, joined by an intermediate plateau-like regime in which the flux is limited by the available amount of O. An important suboxide contribution is expected for all elemental sources whose suboxide exhibits a higher vapor pressure than the element, such as B, Ga, In, La, Si, Ge, Sn, Sb, Mo, Nb, Ru, Ta, V, and W. This contribution can play a decisive role in the molecular beam epitaxy of oxides, including multicomponent or complex oxides, from elemental sources. Finally, our model predicts suboxide-dominated growth in low-pressure chemical vapor deposition of Ga$_{2}$O$_{3}$ and In$_{2}$O$_{3}$.
\end{abstract}

\maketitle

\section{Introduction}
\label{sec:introduction}
Transparent conducting oxides (TCOs), transparent semiconducting
oxides (TSOs), and multicomponent or complex oxides gained more and more interest within the last years because of their great potential for electronic devices, photovoltaics and sensors\cite{wang2016a,ramesh2008,schlom2015,higashiwaki2018,Lorenz2016,Coll2019}.
For the growth of device quality single-crystalline thin films, molecular beam epitaxy (MBE) established itself as the major growth technique \cite{raghavan2016,higashiwaki2012,cheng2021,park2020}, while low pressure chemical vapor deposition (LPCVD) has been demonstrated to be capable of delivering thick films of high quality for the case of  Ga$_2$O$_3$ \cite{rafique2016,feng2019a} and In$_2$O$_3$ \cite{karim2018, zhang2019}.
For the case of MBE, semiconducting oxides (e.g. ZnO \cite{nishimoto2008}, SnO \cite{budde2020}, SnO$_{2}$ \cite{tsai2009a,vogt2015a},
In$_{2}$O$_{3}$ \cite{vogt2015a,bourlange2008}, and Ga$_{2}$O$_{3}$ \cite{vogt2015a,tsai2009,sasaki2012}) as well as complex oxides  have been grown by the reaction of the vapor from a metal charge placed
in a heated effusion cell with reactive oxygen (an oxygen plasma or
ozone) on the heated substrate in an ultra-high vacuum chamber \cite{schlom2015,Engel-Herbert2013}. During LPCVD, metal sources and substrate are essentially placed in a tube furnace held at a temperature ranging from 900 to 1050$^{\circ}$C, and a mixture of Ar and O$_{2}$ with total pressure in the mbar-range passes over the liquid Ga or In to transport Ga- or In-containing species to the substrate, on which they form the oxide film with the O$_2$ \cite{zhang2020,rafique2016,feng2019a,karim2018, zhang2019}. For the case of Ga$_2$O$_3$ growth, the Ga-containing species have been described as Ga evaporating from the liquid Ga source \cite{zhang2020}.

Regarding MBE grown samples, a critical inspection of reported Ga$_{2}$O$_{3}$ growth rates \cite{oshima2017,oshima2018,cheng2018a} and Sn-concentrations in doped Ga$_{2}$O$_{3}$ \cite{ahmadi2017,kracht2017a,han2018a} and In$_{2}$O$_{3}$ \cite{bierwagen2013a}  testifies unexpectedly high metal incorporation into the films for the used, comparably low metal effusion-cell temperature. Indeed, unexpectedly high Si concentrations in MBE grown, Si-doped Ga$_{2}$O$_{3}$ have recently been reported by \citet{kalarickal2019a}, and attributed to the formation and desorption of volatile SiO at the Si effusion cell, that exceeds the expected Si flux.
This additional, unintentional mechanism of SiO formation at comparably
low Si temperatures can be understood as a direct oxidation of the
elemental Si charge in the oxygen background of the growth chamber
and plays a key role for the understanding of Si doping concentrations
in oxide MBE \cite{kalarickal2019a}.

In this article, we show that this mechanism is crucial for understanding metal incorporation in the epitaxy of oxides in general. In particular, we demonstrate experimentally that the metal incorporation is dominated by a contribution  
due to the unintentional formation and evaporation of the suboxide. Specifically, we investigate the metal sources Sn and Ga by quadrupole mass spectrometry (QMS) with respect to their metal and suboxide fluxes when exposed to an oxygen background that is typical for oxide MBE.
In addition, we developed a kinetic
model that allows us to quantitatively describe the total (suboxide and metal) flux from the metal
sources when used in oxygen background. This model allows to explain literature data on Sn doping \cite{bierwagen2013a,ahmadi2017,kracht2017a,han2018a} and Ga$_{2}$O$_{3}$
growth rate \cite{oshima2017,oshima2018,cheng2018a} in oxide MBE, and predicts the suboxide flux from the metal source to be orders of magnitude higher than the pure metal flux during reported growth of Ga$_{2}$O$_{3}$ \cite{rafique2016,feng2019a,zhang2020}
and In$_{2}$O$_{3}$ \cite{karim2018, zhang2019} by LPCVD. Finally, we predict by the comparison of the vapor pressure curves of selected pure elements to those of their suboxide, that dominant suboxide desorption from the elemental source can be relevant for oxide MBE using B, Ga, In, Sb, La, Ge, Si, Sn, Mo, Nb, Ru, Ta, V, and W sources, whereas we can exclude this mechanism for Ba, Al, Ti, and Pb sources.

\begin{figure*} 
\centering
\includegraphics[width=0.45\textwidth]{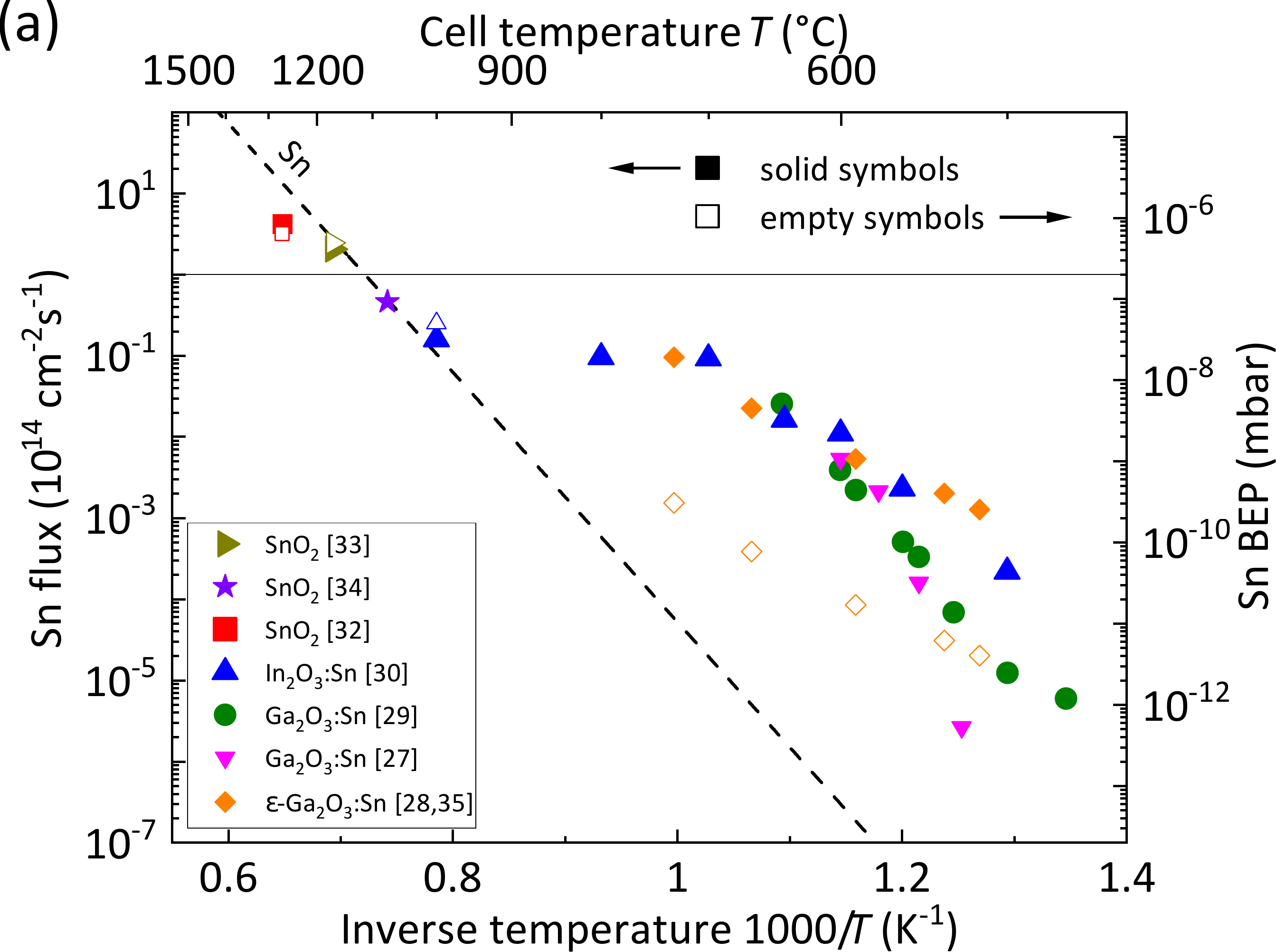}
\qquad{}
\includegraphics[width=0.45\textwidth]{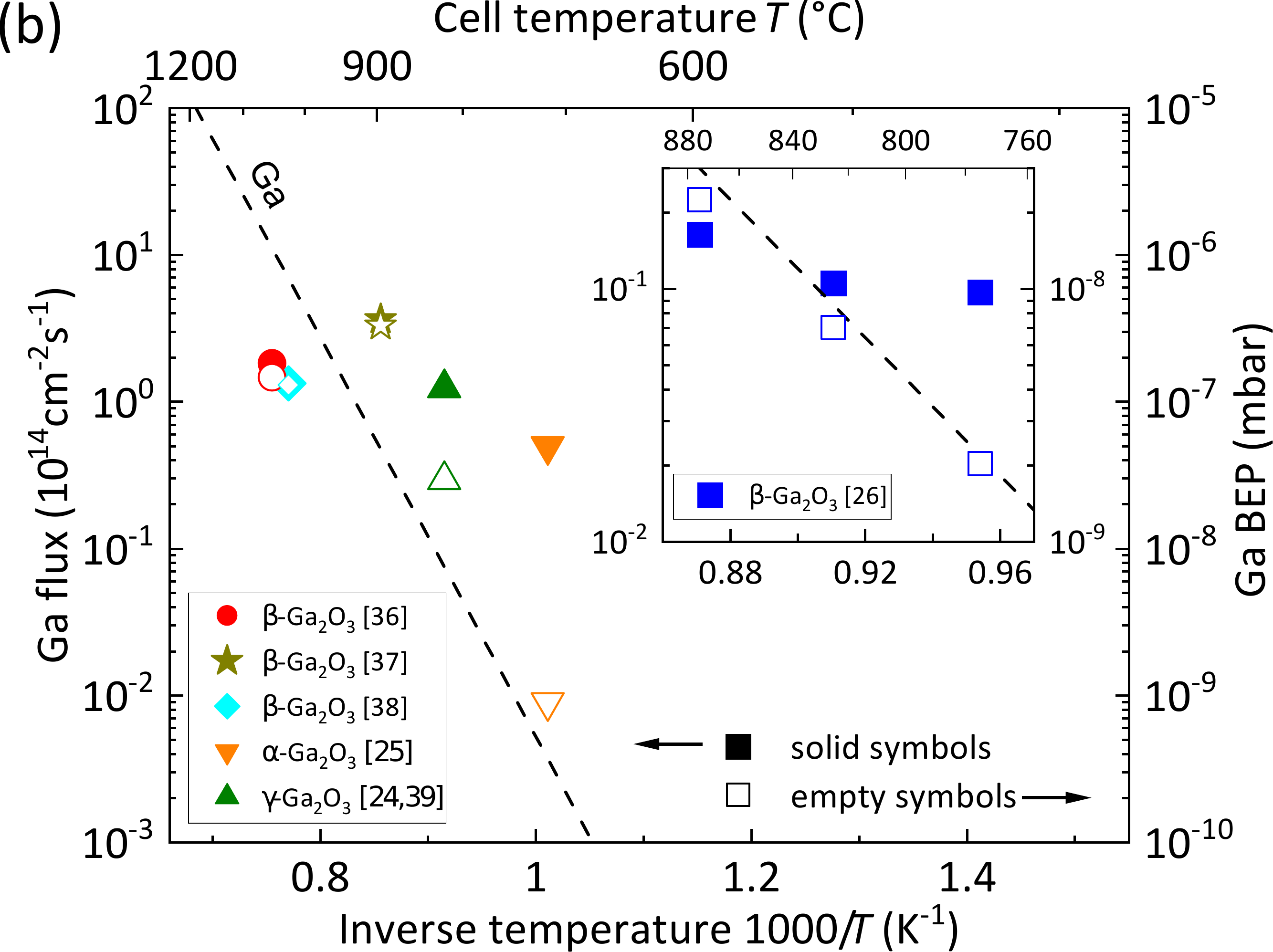}
\caption{Arrhenius diagrams of incorporated metal atoms (left axis, solid symbols) and corresponding BEP (right axis, empty symbols) during MBE-growth.
(a) Incorporated Sn atoms. Data of Refs.~\cite{bierwagen2013a,han2018a,ahmadi2017,white2009a} are based
on layers grown in the same MBE system. Data of Refs.~\cite{simion2019,martinez2018a,kracht2017a,PCEickhoff2020} are based on growth in different MBE systems.
The empty symbols show the reported BEP values of the corresponding fluxes for the data of Refs.~\cite{simion2019,bierwagen2013a,white2009a,kracht2017a} with respect to the right axis. A flux of 10$^{14}$~cm$^{-2}$s$^{-1}$ corresponds to a SnO$_2$-growth rate of $\approx$~2.3~nm/min, and to a BEP of $\approx$~2~$\times$~10$^{-7}$~mbar.
 (b) Incorporated Ga atoms. Data by Refs.~ \cite{vogt2018a,vogt2017a,mazzolini2020, oshima2018,oshima2017,PCOshima2020}, and \cite{cheng2018a}.
A flux of 10$^{14}$~cm$^{-2}$s$^{-1}$ corresponds to a $\beta$-Ga$_2$O$_3$-growth rate of $\approx$~1.6~nm/min, and to a BEP of $\approx$~10$^{-7}$~mbar.\protect 
Note that for (a) and (b) dashed black lines represent the expected metal flux according to the vapor pressure curve extracted from FactSage\texttrademark \cite{factsage}. \label{fig_lit_data}}
\end{figure*}

\section{Reported fluxes from the oxide-MBE literature}
During MBE-growth the metal flux from an effusion cell is essentially proportional to the metal vapor pressure at the given effusion-cell temperature, and it is often measured in the absence of an oxygen background as beam-equivalent pressure (BEP) by a nude ion gauge at the substrate position. Consequently, the dependence of metal flux and BEP on effusion-cell temperature is well described by the vapor-pressure curve of the metal. Figure~\ref{fig_lit_data} summarizes published data on incorporated Sn and Ga fluxes during oxide MBE together with the corresponding measured or extrapolated BEPs for a wide range of metal-effusion-cell temperature  in an Arrhenius diagram with a fixed ratio $\alpha$ between flux- and BEP-scale, and includes dashed lines representing the vapor pressure of the respective metal. Specifically, the incorporated metal fluxes were calculated by multiplying growth rate with doping concentration (of Sn-doped Ga$_{2}$O$_{3}$ \cite{ahmadi2017,kracht2017a,han2018a} and In$_{2}$O$_{3}$ \cite{bierwagen2013a} films) or cation density (of SnO$_2$ \cite{white2009a,martinez2018a,simion2019} or Ga$_{2}$O$_{3}$ \cite{oshima2017,vogt2017a,oshima2018,vogt2018a,mazzolini2020,cheng2018a}  films) for the case of a dopant flux or cation flux, respectively. We assume that the incorporated metal flux reflect the source-metal flux at the substrate since the chosen data largely corresponds to non-metal-rich growth conditions, under which full metal incorporation can be expected \cite{vogt2015a}.

As shown in Fig.~\ref{fig_lit_data}(a), at high Sn-cell temperatures ($T_\text{Sn}\ge 1000~^\circ$C) the incorporated Sn-fluxes and Sn-BEPs (from three different MBE systems) show the expected behavior of approximately following the vapor pressure behavior of Sn, denoted by a dashed line. The ratio of Sn-BEP to Sn-flux is almost equal for the MBE systems used in Refs.~\cite{bierwagen2013a,white2009a} and \cite{simion2019} (BEP and related flux datapoints overlap using the same $\alpha$) suggesting that the BEP can be used to estimate the metal flux independently of the used MBE system. Moving to significantly lower Sn-cell temperature ($T_\text{Sn}\le 800~^\circ$C), however, the incorporated metal flux becomes two to four orders of magnitude higher than the expected metal flux from the Sn-vapor pressure curve or than the respective BEPs. In this low-temperature regime, the incorporated Sn flux also follows an activated behavior up to  $T_\text{Sn}\approx700~^\circ$C and forms a plateau in the transition to the high-temperature regime. The corresponding activation energy is very similar to that of liquid Sn, which is why in Refs.~\cite{han2018a} and \cite{bierwagen2013a} the flux has been attributed to Sn evaporation. Instead, SnO, having the same activation energy, evaporates from the Sn source in the low-temperature regime, as we will show in this article.

A qualitatively similar behavior can be observed for the incorporated Ga-flux shown in Fig.~\ref{fig_lit_data}(b). At high Ga-cell temperature  ($T_\text{Ga}>850~^\circ$C) Ga-BEP and Ga-flux agree well among different MBE systems (see Ref.~\cite{vogt2018a}, and Refs.~\cite{vogt2017a,mazzolini2020}). For the same MBE system the relation of Ga flux to cell temperature can depend strongly on further details [see uncertainty when comparing \cite{vogt2017a} and \cite{mazzolini2020} in Fig.~\ref{fig_lit_data}(b)]. This uncertainty can be related, for example, to different filling levels, cell design, or position of the cell with respect to the substrate.
Nevertheless, data for lower $T_\text{Ga}$, available from other MBE systems (\cite{oshima2018,oshima2017,PCOshima2020} and \cite{cheng2018a}) again show that the incorporated Ga-flux is significantly  higher than expected from the vapor pressure curve as well as measured BEP. This deviation increases with decreasing cell temperature: the BEP decreases according to the Ga-vapor pressure curve, whereas the incorporated flux shows a much weaker dependence on $T_\text{Ga}$, almost forming a plateau [see inset of Fig.~\ref{fig_lit_data}(b)].

Likewise, significantly higher incorporated fluxes than expected from the vapor pressure curve with an almost plateau-like cell-temperature dependence have recently been demonstrated for Si during the MBE growth of Si-doped  Ga$_{2}$O$_{3}$, and explained by the formation and desorption of the volatile suboxide SiO instead of Si from the Si source \cite{kalarickal2019a}.  In the following we will show theoretically and experimentally, that the very same mechanism is responsible for the unexpectedly high Ga- and Sn-incorporation during oxide MBE growth at low cell temperatures. The apparent discrepancy between BEP and incorporated flux can be immediately understood by the absence and presence of an oxygen background pressure (typically in the $10^{-6}$~mbar-range) in the growth chamber during BEP measurement and film growth, respectively.

\section{Experimental}
\begin{figure}[b!]
\centering
\includegraphics[width=0.45\textwidth]{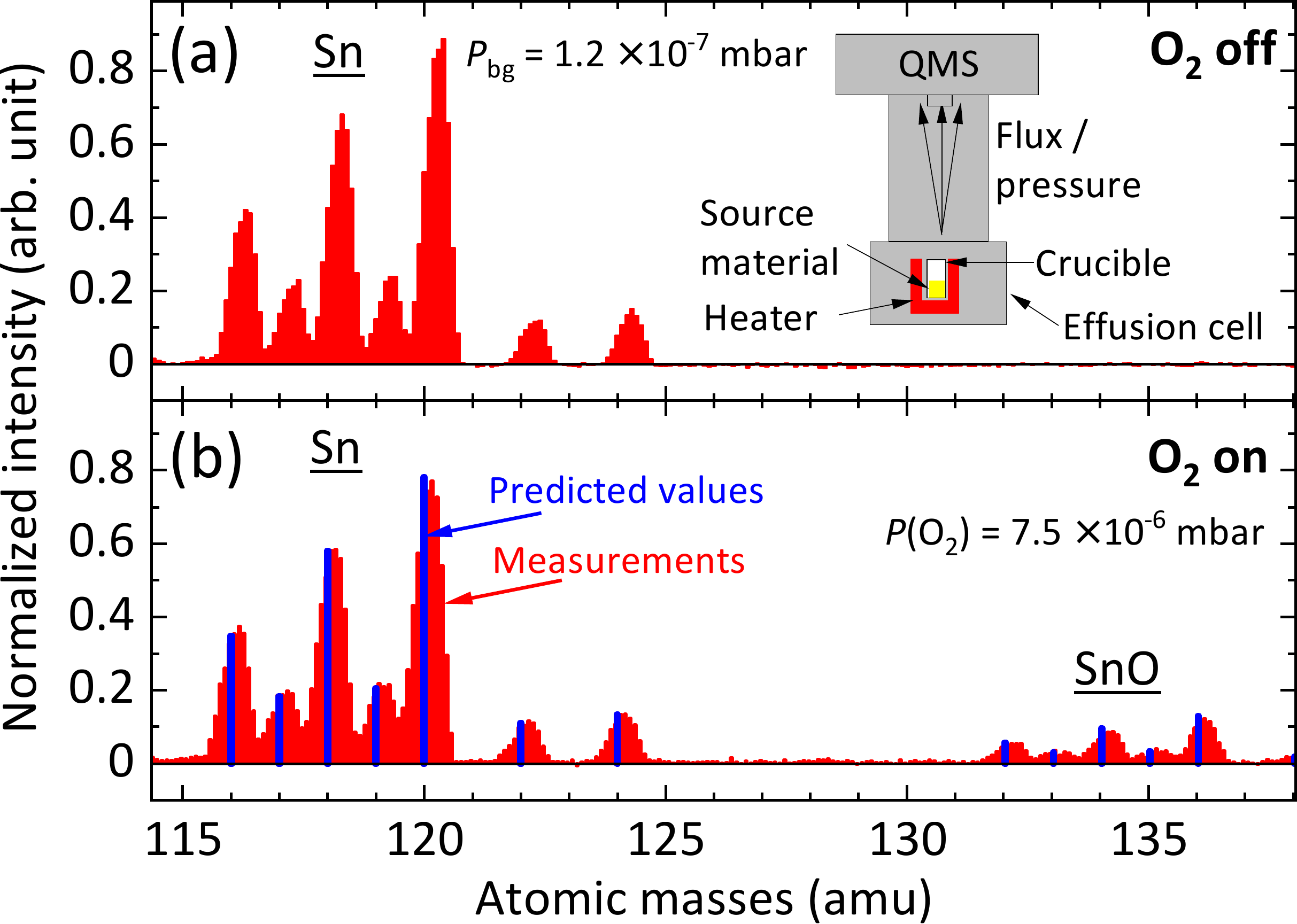}
\caption{(a) QMS spectrum of Sn at Sn cell temperature of 1000$\thinspace ^{\circ}$C when no oxygen is applied. Inset: schematics of experimental setup according to \cite{hoffmann2020a}. (b) QMS spectrum of Sn and SnO recorded at the same conditions as in (a) but with an oxygen pressure of 7.5~$\times$~10$^{-6}$~mbar. The blue lines in (b) are the expected Sn and SnO signals according to their isotopic distribution. The background pressure of the system $P_{\text{bg}}$ is in the low 10$^{-7}$~mbar region.}\label{fig_qms_profile_Sn_SnO}
\end{figure}
Two different methods were used for the experimental investigation. The first method is the measurement of the direct beam flux using a QMS.
The QMS experiments were performed in a test chamber as shown in the inset of Fig.~\ref{fig_qms_profile_Sn_SnO}. Further, Fig.~\ref{fig_qms_profile_Sn_SnO} shows a typical QMS spectrum of the direct Sn flux $j_{Sn}$ and SnO suboxide flux $j_{SnO}$ from a single-filament effusion cell revealing the presence of SnO suboxide that forms in a controlled O-background pressure ($j_O$ ). Due to the isotopic distribution, Sn and SnO can easily be identified.
For the QMS experiments, the device ionizer was run at an electron energy of 50~eV to maximize its sensitivity. As a result, some of the measured signals might be affected by fragmentation \cite{hoffmann2020a}, e.g., by the fragmentation of suboxide molecules into metal and oxygen atoms.  Further details about the used setup and the measurements can be found in Ref.~\citenum{hoffmann2020a}. 

As a second method, we also performed experiments for the Ga$_{2}$O$_{3}$ growth in an MBE system as described in Ref.~\citenum{cheng2018a}. Here, we determined the growth rate by reflection high-energy electron diffraction (RHEED) oscillations  for Ga-cell temperatures ranging from 400 to 875~$^\circ$C. The hot lip of the used dual-filament cell was kept 150~$^\circ$C above the Ga-cell temperature. 
The fluxes $j_\text{cell}$ at the effusion cell (e.g., calculated by the kinetic model) relate to the experimentally obtained ones at the target (QMS or substrate) $j_\text{target}$ through a geometry factor  \cite{franchi_2013}:
\begin{align}
j_\text{target}=\frac{r^{2}}{L^{2}}\text{cos}(\phi)\times j_\text{cell},\label{eq_geometry}
\end{align}
where $r$ is the cell radius, $L$ is the distance between effusion cell and target,
and $\phi$ is the angle of the cell relative to the target. 
For the experimental investigation Sn slugs (5N) and Ga slugs (7N) were used as source charge in the effusion cell.

\section{Kinetic model for suboxide evaporation from an element source}
\begin{figure}[b!]
\includegraphics[width=5cm]{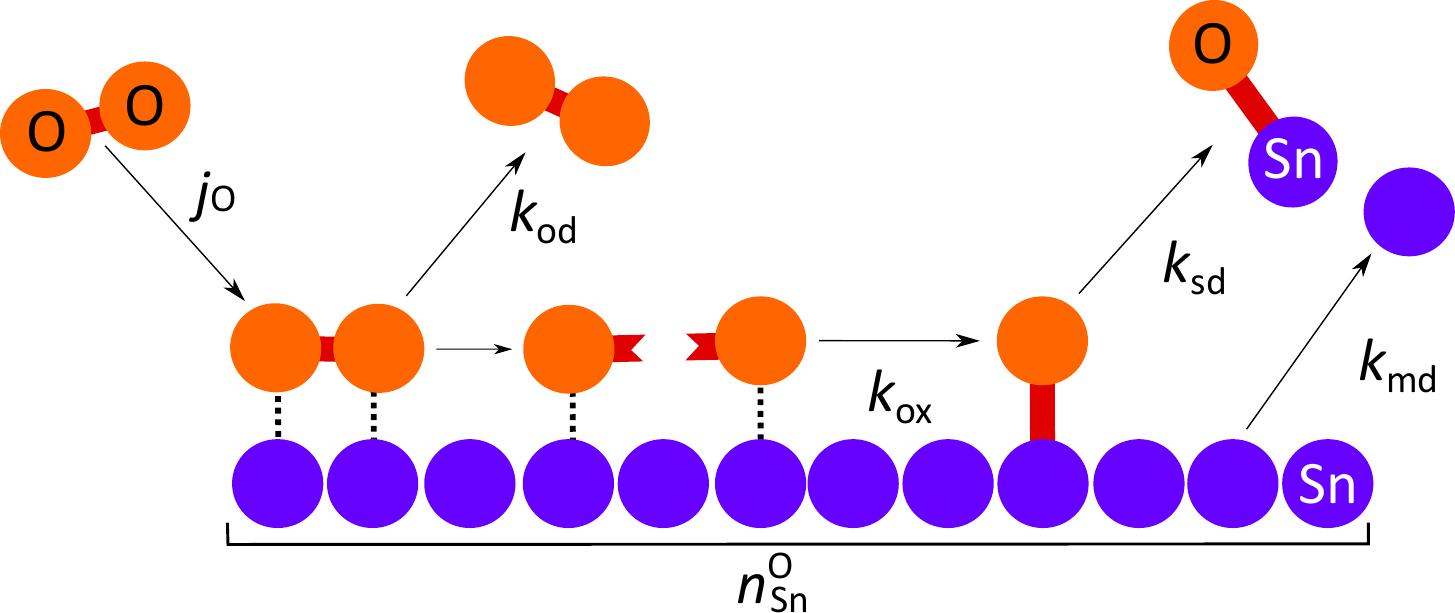}
\caption{Schematic representation of the SnO evaporation from an Sn surface in
O-containing ambient: A flux $j_{\text{O}}$ of O-species (orange discs)
impinges on the Sn (purple discs) surface with $n_{\text{Sn}}^{0}$ sites,
forming a physisorbed (denoted by a dotted vertical line) layer which
can desorb with rate constant $k_{\text{od}}$ or (after a dissociation process) form an SnO molecule (rate
constant $k_{\text{ox}}$) with Sn from the surface. The chemical bonds are
denoted by red lines. Finally, this SnO can desorb with
rate constant $k_{\text{sd}}$ providing the SnO evaporation. In addition, Sn
atoms can desorb from the free Sn surface with rate constant $k_{\text{md}}$.}
\label{fig_subox_schematics}
\end{figure}
As suggested in Ref.~\citenum{kalarickal2019a} the process of
suboxide evaporation from element sources can be described by the sequence
of oxygen adsorption on the element surface (e.g., Si or Sn), reaction
to the suboxide (e.g., SiO or SnO), and desorption of the suboxide
from the surface. Detailed studies on the reaction of molecular oxygen with metallic surfaces Refs.~\citep{lawless1974,panas1989,schmeisser1981} are subdividing the oxidation into physisorption, chemisorption, dissociation, and suboxide formation, as schematically shown in Fig.~\ref{fig_subox_schematics} for the example of SnO formation. In the following, rate-equation-based model, we cannot include these detailed steps due to a lack of experimental access but believe that any barriers related to adsorption/dissociation/oxidation are summarized in the activation for oxidation ($k_{\text{ox}}$). Following Ref.~\citep{schmeisser1981}, that found the reaction of adsorbed oxygen with the Ga surface even below room temperature, however, we expect the dissociation and reaction to be rapid and thus not rate-limiting.

The individual steps in our simplified quantitative model can be described by the following two coupled differential equations
for the concentrations of physisorbed oxygen adatoms $n_{\text{O}}(t)$ and
SnO molecules $n_{\text{SnO}}(t)$: 
\begin{align}
\frac{dn_{\text{O}}(t)}{dt}=j_{\text{O}}(t)-k_{\text{ox}}n_{\text{O}}(t)[n_{\text{Sn}}^{0}-n_{\text{SnO}}(t)]-k_{\text{od}}n_{\text{O}}(t)\label{eq_SnO_dgl_1}\\
\frac{dn_{\text{SnO}}(t)}{dt}=k_{\text{ox}}n_{\text{O}}(t)[n_{\text{Sn}}^{0}-n_{\text{SnO}}(t)]-k_{\text{sd}}n_{\text{SnO}}(t).\label{eq_Sno_dgl_2}
\end{align}

In Eq.~\ref{eq_SnO_dgl_1}, $j_{\text{O}}$ is the impinging oxygen flux,
the second term describes the formation of SnO with the rate constant
$k_{\text{ox}}$ and total number of Sn sites per unit area
$n_{\text{Sn}}^{0}$ out of which $n_{\text{SnO}}(t)$ sites are occupied by SnO
molecules. The third term represents the amount of physisorbed oxygen that desorbs
from the surface with the rate constant $k_{\text{od}}$. Further, in Eq.~\ref{eq_Sno_dgl_2}
the first term describes the SnO formation (same as in Eq.\,~\ref{eq_SnO_dgl_1})
and the second term, $k_{\text{sd}}n_{\text{SnO}}(t)$, describes the desorbing
SnO flux $j_{\text{SnO}}$ from the surface with the rate constant $k_{\text{sd}}$.
In steady state, these coupled differential equations for the time dependent surface coverages $n_{\text{O}}(t)$ and $n_{\text{SnO}}(t)$ reduce to coupled algebraic equations for $n_{\text{O}}$ and $n_{\text{SnO}}$.
Their solution yields a suboxide flux
\begin{align}
j_{\text{SnO}} & =\frac{\alpha}{2}\left(1-\sqrt{1-(\beta/\alpha)^{2}}\right)\label{eq_SnO_flux_sol}\\
\text{with}\quad\alpha & =j_{\text{O}}+\frac{k_{\text{od}}k_{\text{sd}}}{k_{\text{ox}}}+k_{\text{sd}}n_{\text{Sn}}^{0}\nonumber \\
\text{and}\quad\beta & =\sqrt{4j_{\text{O}}k_{\text{sd}}n_{\text{Sn}}^{0}}.\nonumber 
\end{align}
Considering in addition the Sn flux $j_{\text{Sn}}=k_{md}(n_{\text{Sn}}^{0}-n_{\text{SnO}})$ with
the rate constant $k_{\text{md}}$ for metal desorption, the total flux of Sn-containing species $j_{\text{T}}=j_{\text{Sn}}+j_{\text{SnO}}$
is given by 
\begin{align}
j_{T} & =\frac{1}{2}\biggl[\alpha(1-\gamma)\left(1-\sqrt{1-(\beta/\alpha)^{2}}\right)+\frac{\beta^{2}\gamma}{2j_{\text{O}}}\biggr]\label{eq_total_flux_sol}\\
\text{with}\quad\gamma & =\frac{k_{\text{md}}}{k_{\text{sd}}}.
\end{align} The influence of cell temperature on the fluxes in our kinetic model enters through the
rate constants $k_{\text{od}},\thinspace k_{\text{ox}},\thinspace k_{\text{sd}},$ and $k_{\text{md}}$, that are
thermally activated {[}$k_{\text{x}}=A_x\exp(-\frac{E_{\text{x}}}{k_{\text{B}}T})$
with the pre-exponential factor $A_{\text{x}}$, activation energy $E_{\text{x}}$, and cell temperature $T${]}. While $ j_{T}$ contributes to film growth in the presence of an oxygen background (impinging O-flux $j_{\text{O}}>0$), the BEP measurement performed without oxygen background ($j_{\text{O}}=0$) corresponds to $j_{\text{Sn}}$ in the absence of a suboxide coverage ($n_{\text{SnO}}=0$). The impinging O-flux $j_{\text{O}}$ is related to the background oxygen partial pressure $p_{\text{O}_{\text{2}}}$ by kinetic gas theory \cite{franchi_2013}: 
\begin{align}
j_{\text{O}}=2 p_{\text{O}_{\text{2}}}\left(\frac{N_{\text{A}}}{2\pi mk_{\text{B}}T}\right)^{1/2}\label{eq_kinetic_gas}
\end{align}
with the Avogadro constant $N_{\text{A}}$, the molecule mass $m$, Boltzmann constant $k_{\text{B}}$, and oxygen temperature $T$. We use the same equation (without the leading factor 2 taking into account the stoichiometry of the O$_2$ molecule) to convert theoretical vapor pressure data into the corresponding vapor-pressure flux at the effusion cell.  

For the Ga$_{2}$O suboxide desorption, we proceeded in the same manner
as for the SnO desorption. The algebraic steady state equations are
given by the following expressions: 
\begin{align}
0=j_{\text{O}}-k_{\text{ox}}n_{\text{O}}(n_{\text{Ga}}^{0}-n_{\text{Ga}_{2}\text{O}})^{2}-k_{\text{od}}n_{\text{O}}\label{eq_Ga2O_steady_1}\\
0=k_{\text{ox}}n_{\text{O}}(n_{\text{Ga}}^{\text{0}}-n_{\text{Ga}_{2}\text{O}})^{2}-k_{\text{sd}}n_{\text{Ga}_{2}\text{O}}.\label{eq_Ga2o_steady_2}
\end{align}
The solution of the resulting cubic equation is straightforward, but too bulky to be given here explicitly.

\section{Results and Discussion}

\subsection{Influence of effusion cell temperature at constant oxygen background pressure}
\begin{figure*}[t!]
\centering \includegraphics[width=0.43\textwidth]{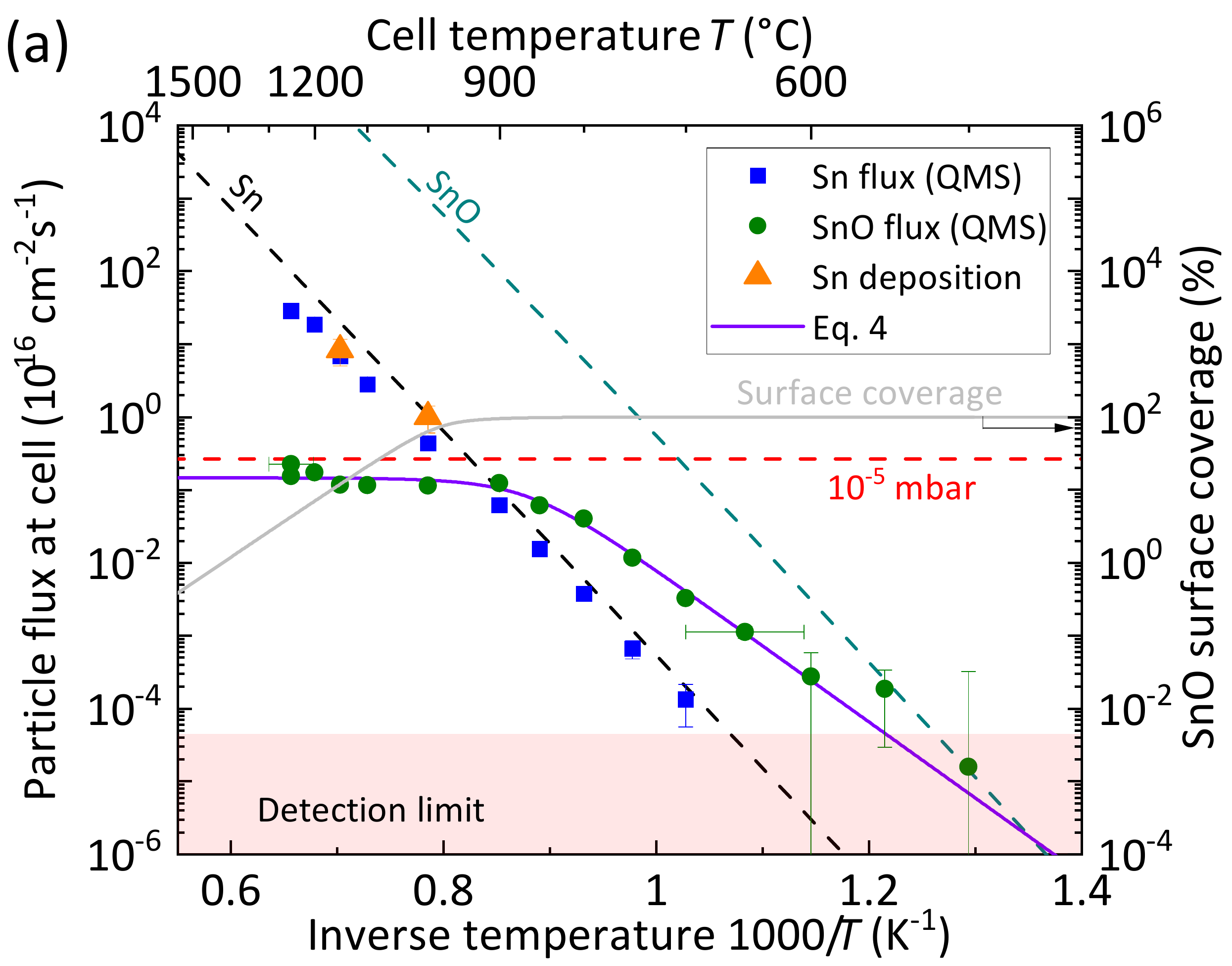}\qquad{}\includegraphics[width=0.43\textwidth]{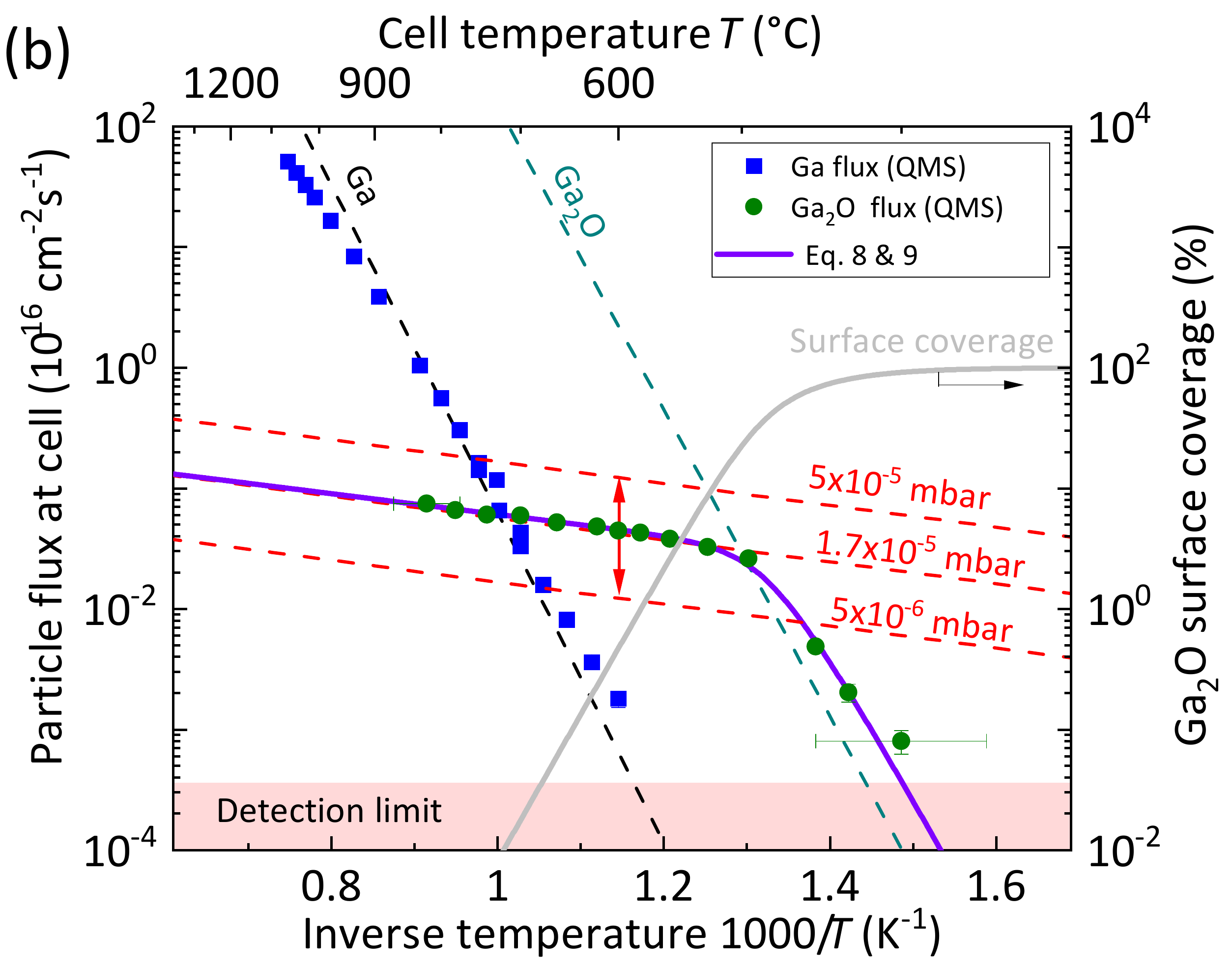}
\caption{Arrhenius diagram of Sn and SnO (a) and Ga and Ga$_{2}$O (b) fluxes
calculated from QMS experiments using Eqs.~\ref{eq_kinetic_gas} and \ref{eq_geometry}.
For reference, we show the expected Sn and SnO (a) as well as Ga and Ga$_2$O (b) fluxes (black and cyan dashed lines) based on their vapor pressure curves. The suboxide flux (purple solid line) was fitted
according to (a) Eq.~\ref{eq_SnO_flux_sol}~, and (b) to the solution
of Eqs.~\ref{eq_Ga2O_steady_1} and \ref{eq_Ga2o_steady_2}. The grey solid line describes the relative surface coverage with the suboxide ($n_{\text{SnO}}/n_{\text{Sn}}^0$ and  $n_{\text{Ga}_{2}\text{O}}/n_{\text{Ga}}^0$ for the steady-state solutions of Eqs.~\ref{eq_SnO_dgl_1}, \ref{eq_Sno_dgl_2} and Eqs.~\ref{eq_Ga2O_steady_1},~\ref{eq_Ga2o_steady_2}) in relation to the right axis. 
The red dashed line in (a) denotes the oxygen flux at 10$^{-5}$~mbar. In (b) the red dashed lines depict the given oxygen flux of 0.5,~1.7,~and~5~$\times 10^{-5}$ mbar in addition to an
activation energy of 0.18~eV.
In (a) and (b), horizontal error bars of exemplary suboxide data points the influence of a possible
systematic error of 50$\thinspace^{\circ}$C in the high and low temperature
regime.}
\label{fig_sn_sno_model_data_qms}
\end{figure*}
The influence of the effusion cell temperature at a constant oxygen background pressure
on the evaporation of SnO and Ga$_{2}$O from Sn and Ga cells,
respectively, was studied and compared to the corresponding metal
evaporation in the absence of oxygen. 
In Figs.~\ref{fig_sn_sno_model_data_qms}(a) and ~\ref{fig_sn_sno_model_data_qms}(b), the SnO and Ga$_{2}$O fluxes at the effusion cell in an oxygen background of
$p_{\text{O}_2}=10^{-5}$~mbar and the corresponding metal fluxes without
oxygen supply ($p_{\text{O}_2}<10^{-7}$mbar) are shown. 
The QMS signals were converted into fluxes taking into account molecular masses, isotopic distribution, ionization cross section, and geometry (Eq.~\ref{eq_geometry}) factor.
For reference, $p_{\text{O}_2}$ as well as the corresponding theoretical metal and suboxide
vapour pressures $P_{\text{vap}}$ (taken from FactSage\texttrademark \cite{factsage}) were converted into fluxes $J$
using Eq.~\ref{eq_kinetic_gas}, and are indicated by the red as well as black and cyan dashed lines, respectively, which we will label as "vapor pressure" values.
Note, that we added in Fig.~\ref{fig_sn_sno_model_data_qms}(b) a slope to the oxygen partial pressure (red dashed
line) taking into account the thermally activated rate constants $k_{od}$ and $k_{ox}$ as discussed in the theoretical discussion Eq.~\ref{eq_SnO_flux_sol}.
The good agreement of the measured metal fluxes [blue solid
squares in Figs.~\ref{fig_sn_sno_model_data_qms}(a) and ~\ref{fig_sn_sno_model_data_qms}(b)] to the
ones derived from the metal vapor pressure curves (dashed black lines) whose activation energy values are listed in Tab.~\ref{tab_qms_data}, validates the used geometry factor for Eq.~\ref{eq_geometry}.

\begin{table}[b!]
\caption{Activation energies (eV) of the different branches in Figs.~\ref{fig_sn_sno_model_data_qms}(a) and (b). Activation energies of Sn and Ga metal evaporation were calculated using the full QMS dataset.}\label{tab_qms_data}
\begin{ruledtabular}
\begin{tabular}{c c c c}

• &Kinetic model& QMS  & Vapor pressure \\ 
\hline 
Sn & &  2.82 $\pm$ 0.04 & 3.1 \\ 

SnO & 1.9 $\pm$ 0.3 &  & 3.1 \\ 

SnO plateau & 0.19 $\pm$ 0.17 &  & --- \\ 

Ga &  & 2.31 $\pm$ 0.05 & 2.7 \\ 

Ga$_2$O & 2.3 $\pm$ 0.1 &  & 2.3 \\ 

Ga$_2$O plateau & 0.18 $\pm$ 0.02 &  & --- \\ 

\end{tabular} 
\end{ruledtabular}
\end{table}

Next, we apply Eq.~\ref{eq_SnO_flux_sol} [purple solid line in Fig.~\ref{fig_sn_sno_model_data_qms}(a)~and~\ref{fig_sn_sno_model_data_qms}(b)] to the measured suboxide data (green circles) in the presence of an oxygen background. We can identify two regimes for the suboxide flux: a low-temperature regime, where the suboxide flux is following the suboxide vapor pressure curve, and a high-temperature regime where the suboxide flux forms a plateau that is limited by the amount of supplied oxygen required for suboxide formation. The transition temperature between these two regimes  is $\approx550\thinspace^{\circ}$C, and $\approx900\thinspace^{\circ}$C for Ga$_2$O and SnO, respectively. For both branches, the activation energies are also listed in Tab.~\ref{tab_qms_data}. 

As expected, the activation energy in the low-temperature branch agrees well with that of the suboxide vapor pressure in the case of Ga$_2$O, whereas a significant discrepancy of unclear origin is apparent for SnO. Neither the consideration of the higher oligomers Sn$_{2}$O$_{2}$ and Sn$_{4}$O$_{4}$ that contribute to the total suboxide flux which is comprised in the theoretic calculations nor the assumption of a systematic error in the temperature measurement at low temperatures can account for this deviation.
The plateau in the high-temperature regime also shows an activated behavior with significantly lower activation energy of $E_{od}-E_{ox}$ that is related to the $k_{od}/k_{ox}$ term in the definition of $\alpha$ in Eq.~\ref{eq_SnO_flux_sol}. We illustrate this activation in Fig.~\ref{fig_sn_sno_model_data_qms}(b) by an activation energy of 0.18~eV for the O$_2$-flux according to the calculated value of Tab.~\ref{tab_qms_data} for the Ga$_2$O-plateau. 
Additionally added red dashed lines in Fig.~\ref{fig_sn_sno_model_data_qms}(b) illustrate the expected shift of the Ga$_2$O suboxide plateau when the system is exposed to other oxygen fluxes. With increasing (decreasing) oxygen flux, the plateau moves to higher (lower) temperatures and suboxide fluxes.

Our model further provides information about the relative coverage
of the metal surface by its suboxide when exposed to an oxygen background.
The grey, solid lines in Fig.~\ref{fig_sn_sno_model_data_qms}(a) and (b) show with respect to the
right axis that at low temperatures (e.g., standby temperatures of
metal cells in oxide-MBE) the Sn and Ga surfaces can be assumed to be completely
covered by their suboxide. At higher temperature on the plateau of the suboxide flux, the relative coverage decreases exponentially
with increasing temperature, opening up surface area for metal desorption. Additional experiments showed that the Ga surface at
300$^{\circ}$~C can be fully covered by Ga$_{2}$O within $\approx$~30~minutes,
when exposed to an oxygen pressure of 10$^{-5}$~mbar. Our experiments further revealed that
a heating of the Ga cell up to 800$^{\circ}\thinspace$C in the absence
of oxygen is sufficient in order to mitigate this memory effect. Note that we performed our measurements in the test chamber by heating the cell above 800$^{\circ}\thinspace$C in the absence of oxygen before we went to the desired temperature and applied the oxygen pressure to the system. We observed a similar behavior of the Sn cell, and acted in the same manner as for the Ga cell but with a critical temperature of 1100$^{\circ}\thinspace$C
In conclusion, our findings show that suboxide evaporation from metal sources is an even more complex phenomenon than described by our kinetic steady-state model.

As a consequence of our findings, the total (metal+suboxide) flux $j_T$ in steady state that will contribute to film growth is expected to follow an $\textsf{S}$-shaped curve with three different branches, that are well resolved in the case of Ga [Fig.~\ref{fig_sn_sno_model_data_qms}(b)]: at the high-temperature branch $j_T$ follows the metal vapor pressure curve and consists mainly of Ga atoms, on the plateau at intermediate temperatures $j_T$ is mainly given by 
a saturated suboxide flux that is transport-limited by the incoming oxygen flux from the oxygen background pressure,
and at the low-temperature branch the desorption-limited suboxide flux follows the suboxide vapor pressure curve.
 The impact of our findings on film growth will be discussed in the following sections.

\subsection{Role of oxygen background pressure and application to LPCVD}\label{subsec_ox_role}
\begin{figure*}[t!]
\includegraphics[width=0.45\textwidth]{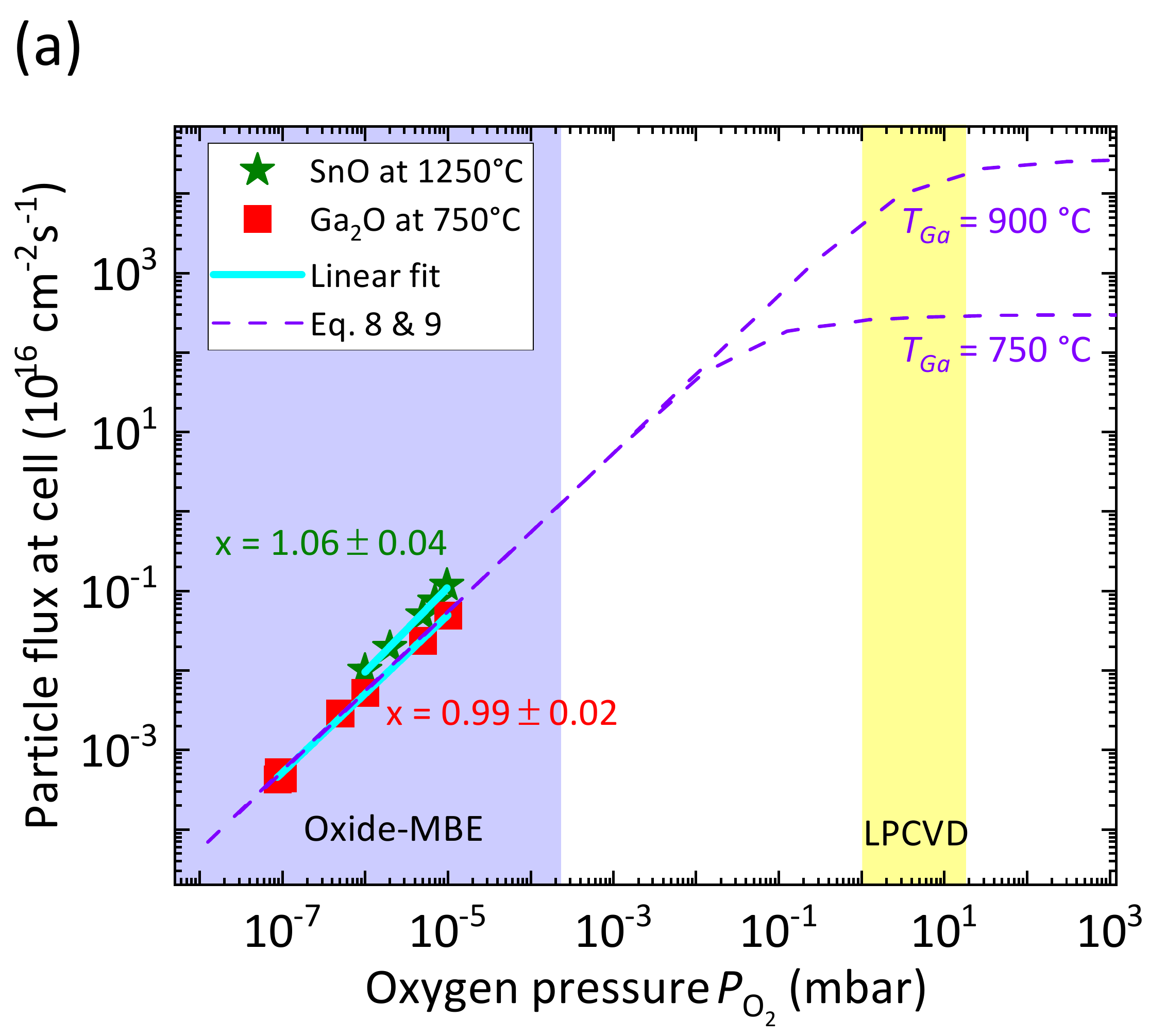}
\includegraphics[width=0.45\textwidth]{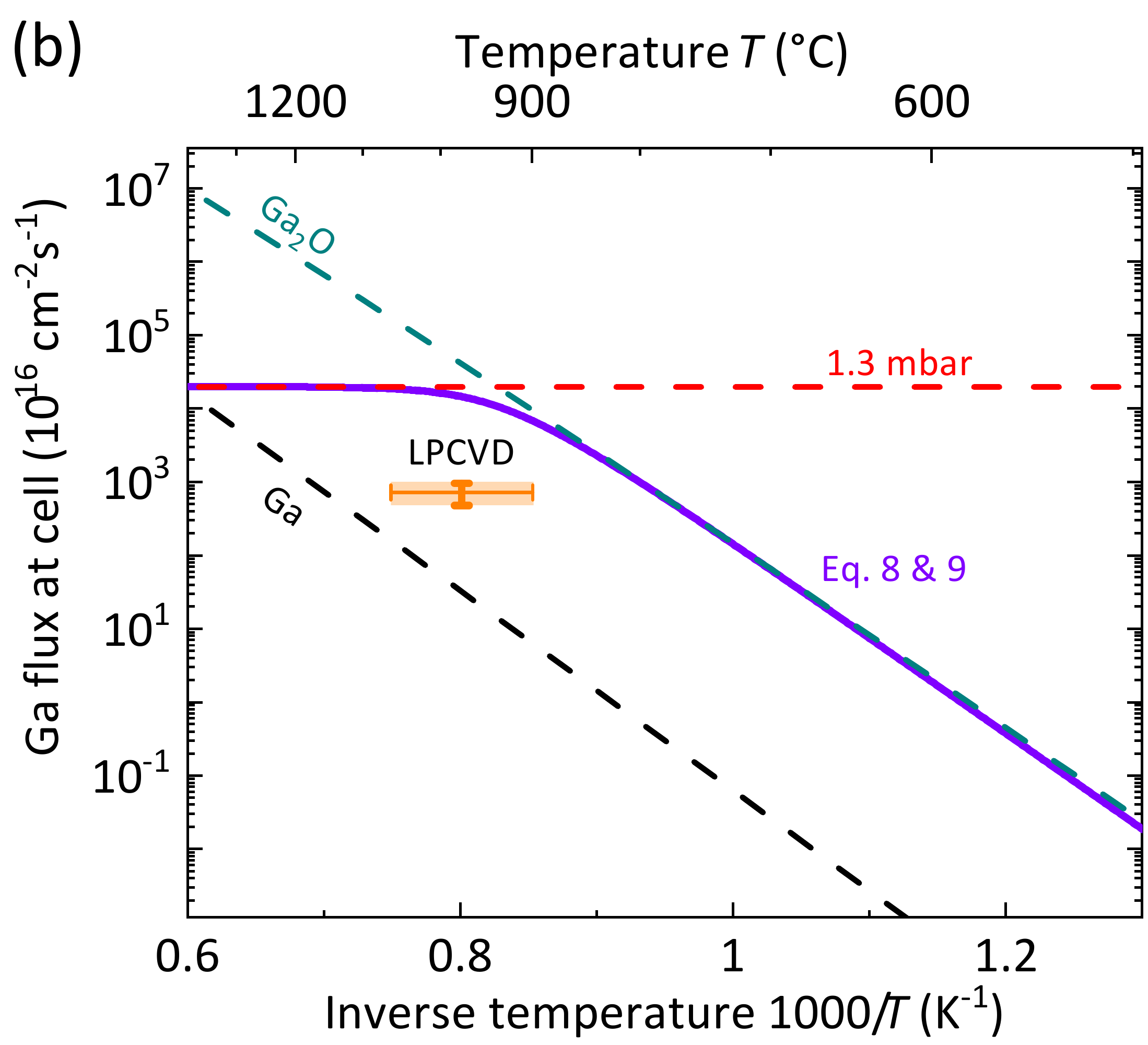}
\caption{(a) Suboxide flux at cell as a function of oxygen background pressure for Sn cell at 1250$\thinspace^{\circ}$C and Ga cell at 750$\thinspace^{\circ}$C.
The solid bright blue lines denote fits according
to the power law relation $y\propto a^{x}$ with x~=~1.06 and 0.99 for SnO and Ga$_2$O, respectively. The dashed purple line denotes the expected suboxide behavior according to Eqs.~\ref{eq_Ga2O_steady_1} and~\ref{eq_Ga2o_steady_2}.
(b) Ga and Ga$_2$O flux at the metal source as function of inverse cell temperature for LPCVD regime. The black and cyan dashed lines denote the theoretically calculated Ga and Ga$_2$O suboxide fluxes. The orange data point is an estimated Ga flux window (orange shaded area) from Ga mass loss of the Ga target according to \cite{PCHongping}. The purple solid line denotes the expected Ga$_2$O flux according to our model assuming an oxygen flux of 1.3~mbar (red dashed line) that is typically used for LPCVD  \cite{zhang2020}.}
\label{fig_subox_vs_O2}
\end{figure*}
Next, we investigate the impact of the oxygen background pressure on the formation of suboxide vapour under conditions suitable for PA-MBE, and compare our model in a second step with published LPCVD data at significantly higher oxygen pressures. Figure~\ref{fig_subox_vs_O2}(a) shows the suboxide flux from the pure
metal charges as function of externally supplied oxygen background
pressure at temperatures that are well into the plateau-regime of suboxide evaporation.
The observed linear dependence of the oxygen flux on the oxygen background confirms that at these
temperatures the suboxide formation is oxygen transport limited, which is also reflected by the kinetic model (see purple dashed line). The oxygen-transport limited regime of SnO evaporation qualitatively explains the increasing Sn-concentration with O/In-BEP ratio reported for Sn-doped In$_2$O$_3$ films reported in Ref.~\citenum{bierwagen2013a}.

Moving to oxygen partial pressures in the mbar-range that are used in LPCVD of Ga$_2$O$_3$ and indicated by the yellow area \cite{feng2019a,zhang2020}, this linearity results in orders of magnitude higher Ga$_2$O fluxes at the source. The saturation at high pressures indicates the suboxide-desorption limited regime for two exemplarily shown Ga temperatures. Figure~\ref{fig_subox_vs_O2}(b) shows the modeled Ga$_2$O and Ga fluxes at an oxygen partial pressure of $\approx$1.3~mbar as a function of source temperature. For comparison, the orange data point including its error bars reflects the Ga flux at the source that corresponds to the observed Ga consumption during LPCVD growth (0.5 to 1~g/hour at 900--1050$^\circ$C at an available Ga surface area of $\approx0.25$~cm$^2$ \cite{PCHongping}). This flux is approximately one order of magnitude larger than the theoretical Ga flux indicating that Ga$_2$O$_3$ growth in LPCVD cannot be explained by Ga evaporation from the source. Ga$_2$O evaporation dominates instead, however, at a lower flux than predicted from the kinetic model (purple line). We attribute this deviation to the fact that the model was developed for a ballistic transport regime whereas LPCVD operates in a diffusive transport regime.

Consequently, the suboxide evaporation of MBE sources depends linearly on the background oxygen pressure when the source temperature is in the plateau regime. The LPCVD growth of Ga$_2$O$_3$ is related to the formation and evaporation of Ga$_2$O from the Ga source that is transported to the substrate. Accoring to the kinetic model, higher Ga$_2$O$_3$ growth rates in the LPCVD can be achieved by increasing the Ga$_2$O flux through higher oxygen partial pressure (lifting the plateau) or a higher growth temperature (to enable maximum suboxide desorption) from the Ga source.

\subsection{Impact of metal suboxide formation on oxide MBE growth}
\begin{figure*}[t!]
\centering 
\includegraphics[width=0.45\textwidth]{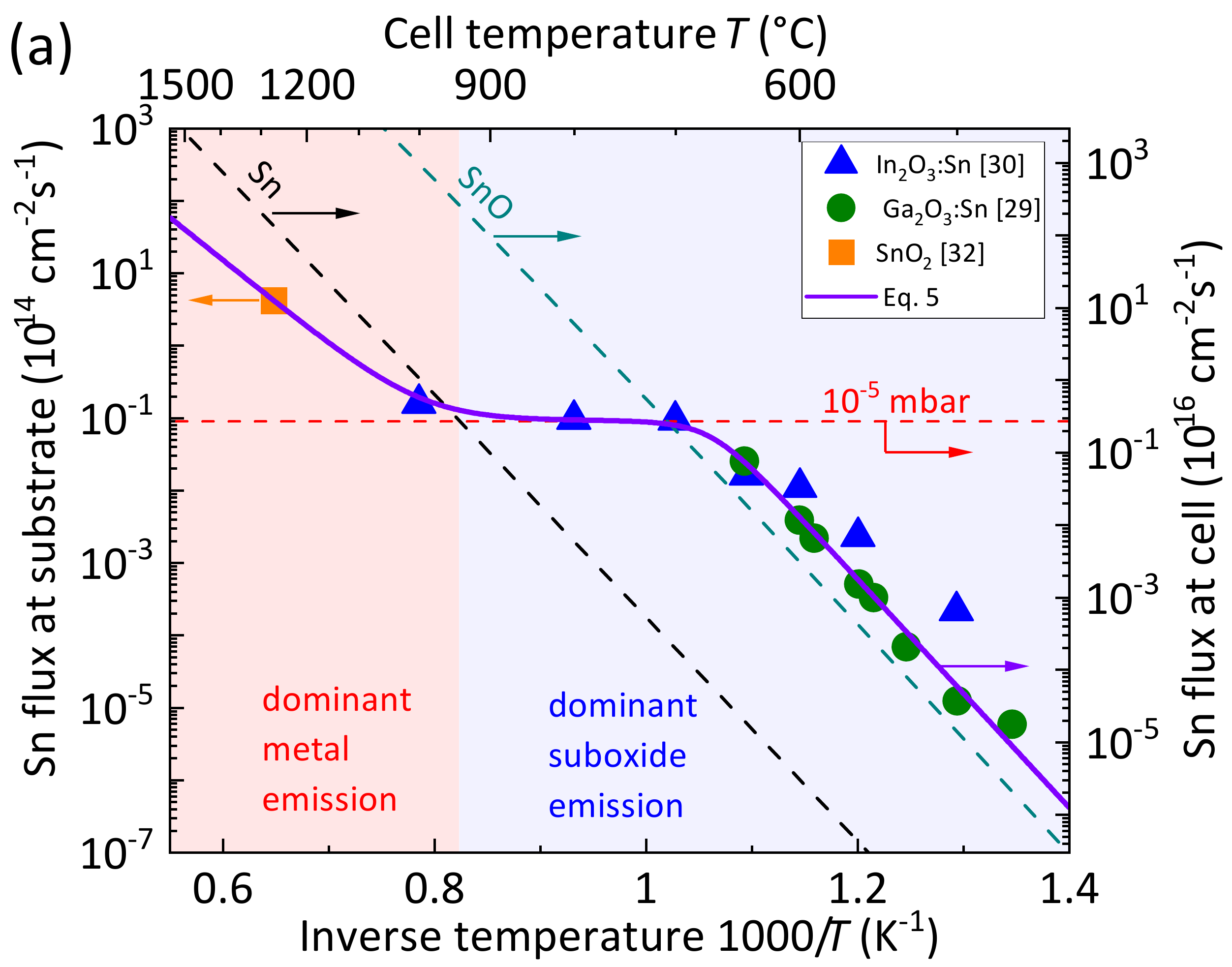}
\includegraphics[width=0.45\textwidth]{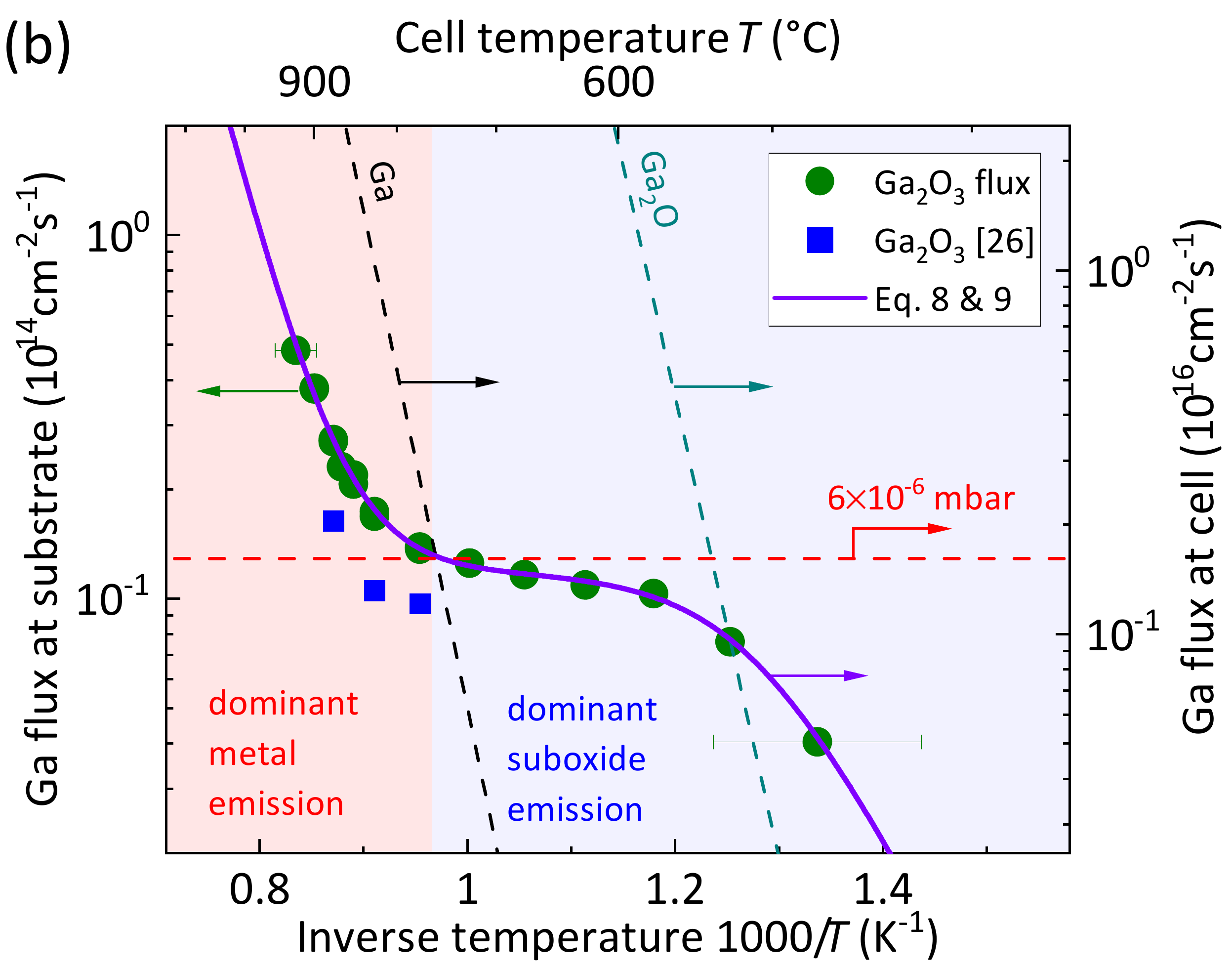}
\caption{Arrhenius diagrams of incorporated cations in MBE growth (symbols) in comparison to the total flux predicted by the kinetic model (solid purple line). The red dashed line illustrates the O$_{2}$ flux. Regions of dominant metal evaporation (red shaded) and suboxide evaporation (blue shaded) are defined according to the crossing of metal flux (dashed black line) and oxygen-limited suboxide plateau (dashed red line). 
(a) Incorporated total Sn flux. The total
flux according to Eq.~\ref{eq_total_flux_sol} was fitted to the data of Refs.~\cite{white2009a,bierwagen2013a,han2018a}, all of which used the same MBE system. For the data points \citet{bierwagen2013a} the oxygen background pressure was $10^{-5}$~mbar. (b) Incorporated total Ga flux determined from Ga$_2$O$_3$ growth rates based on RHEED oscillations at different Ga cell temperatures at an oxygen background pressure of $6\times10^{-6}$~mbar. In addition the data of Ref.~\cite{cheng2018a} are shown which were obtained in the same MBE system.
The purple line describes the total growth rate of Ga$_{2}$O$_{3}$ based on the solution of Eqs.~\ref{eq_Ga2O_steady_1}~and~\ref{eq_Ga2o_steady_2}.\label{fig_SnO_doping_sims_Ga2O3_growth}}
\end{figure*}

\begin{table}[b]
\caption{Activation energies (eV) of metal and suboxide branches shown in Figs.~\ref{fig_SnO_doping_sims_Ga2O3_growth}(a) and \ref{fig_SnO_doping_sims_Ga2O3_growth}(b) calculated from our model and compared to vapor pressure values}\label{tab_sims_rheed_data_model}
\begin{ruledtabular}
\begin{tabular}{c c c}

• & Kinetic model & Vapor pressure \\ 
\hline 
Sn & 2.3 $\pm$ 0.2 & 3.1 \\ 

SnO & 3.12 $\pm$ 0.01 & 3.1 \\ 
 
Ga & 2.2 $\pm$ 0.1 & 2.7 \\ 

Ga$_2$O & 1.2 $\pm$ 0.2 & 2.3 \\ 

\end{tabular} 
\end{ruledtabular}
\end{table}
To show the relevance of our findings on oxide MBE growth we compare the total (metal + combined metal and suboxide) cation flux predicted by the kinetic model to published data on Sn incorporation from recent oxide MBE-growth studies as well as data on Ga incorporation of our own Ga$_2$O$_3$ MBE growth experiment. The related Arrhenius diagrams, Figs.~\ref{fig_SnO_doping_sims_Ga2O3_growth}(a)~and~\ref{fig_SnO_doping_sims_Ga2O3_growth}(b), clearly reflect the $\textsf{S}$-shaped curve predicted by the kinetic model for the total cation flux for both Sn and Ga. In order to take the geometry factor (Eq.~\ref{eq_geometry}) into account, the measured data is shown with respect to the left axis, whereas all solid and dashed lines are given with respect to the right axis reflecting the situation at the metal cell. The derived activation energies are listed in Tab.~\ref{tab_sims_rheed_data_model}. For the case of Sn/SnO, the activation energy of 3.13~eV in the low
temperature regime matches very well with the theoretical one (dashed
cyan line) calculated for SnO vapour pressure
of 3.07~eV. In the high temperature regime, the slight difference
between the activation energies of our model of 2.3~eV and the theoretical
one (dashed black line) of 3.1~eV can be attributed to the lack of
Sn data points. 

We observe the same qualitative behaviour for the Ga$_{2}$O$_{3}$  growth which is shown in
Fig.~\ref{fig_SnO_doping_sims_Ga2O3_growth}(b). The Ga-containing flux incorporated during growth matches very
well with our model in the high and mid temperature regime. 
The activation energy for the Ga branch is 2.2~eV and thereby close
to the Ga activation energy of 2.7 eV calculated from the vapor pressure
(dashed black line).
In the low temperature regime, a larger difference between the activation
energies of our model (1.2~eV) and the theoretically calculated one
{[}2.5 eV (dashed cyan line){]} can be attributed to the lack of data
points at lower cell temperatures. Further, a systematic error caused
by differences in the measured and real cell temperatures at lower
temperatures could be another reason for deviations of the measured
values from the theoretically expected ones {[}dashed cyan lines in
Figs.~\ref{fig_SnO_doping_sims_Ga2O3_growth}(a), and \ref{fig_SnO_doping_sims_Ga2O3_growth}(b){]}. In addition,
the influence of a hot-lip cell on the suboxide
flux at the lower temperatures might also be different from the one
at higher temperatures, i.e. still working as suboxide reservoir.

The good fit of the incorporated cation flux to our kinetic model for the source is consistent with an essentially constant incorporation probability (most likely full incorporation) of the impinging cation flux into the film, as expected for oxygen-rich growth conditions. Our previous models of the incorporation kinetics on the growth surface in oxide MBE growth from the metal flux \cite{vogt2015a,vogt2018b} or suboxide flux \cite{vogt2021} predict with increasing source temperature (and thus source flux) an increasing oxide growth rate (full incorporation in the oxygen-rich regime) followed by a constant growth rate (suboxide-rich regime in the case of a suboxide flux) or a decreasing growth rate (metal-rich regime in the case of a metal flux). Thus, the further increasing incorporated flux following the plateaus in Figs.~\ref{fig_SnO_doping_sims_Ga2O3_growth}(a)~and~(b) cannot be explained by incorporation kinetics but only by our kinetic model for the source.


\subsection{Prediction of dominant suboxide evaporation in oxygen background for different elements}

\begin{figure}[t!]
\centering
\includegraphics[width=0.45\textwidth]{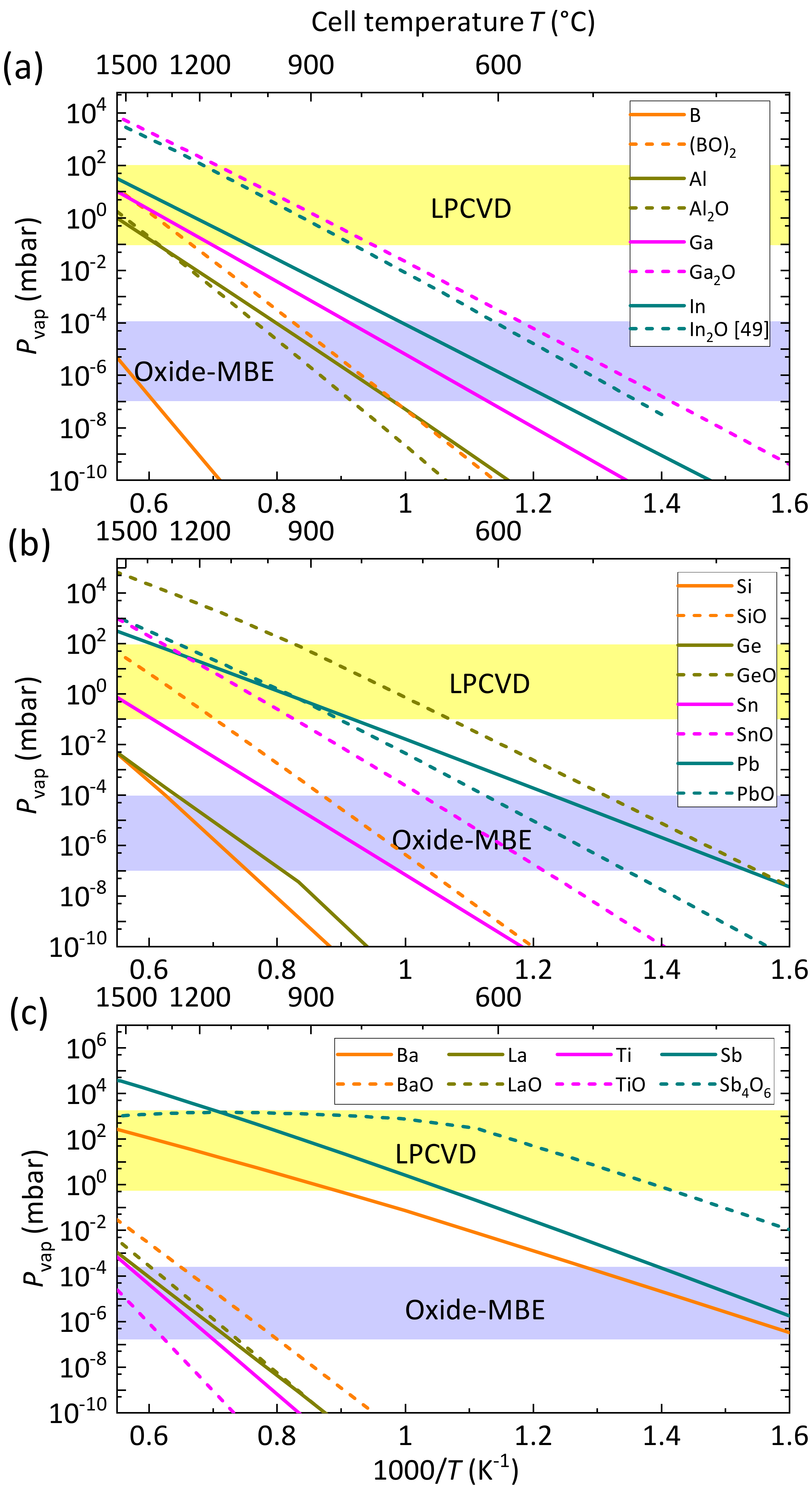}
\caption{(a)-(c) Arrhenius diagrams of calculated vapor pressure curves of elements (solid line) and their corresponding suboxides (dashed line) using FactSage\texttrademark \cite{factsage}. The vapor pressure curve of In$_2$O was taken from Ref.~\citenum{adkison2020}. The blue shaded area denotes the pressure commonly used for oxide growth whereas the yellow shaded area denotes the oxide growth regime in LPCVD. }\label{fig_vap_pressure_all}
\end{figure}

A necessary precondition for dominant suboxide evaporation of elemental sources in an oxygen background is a higher vapor pressure $P_{\text{vap}}$ of the suboxide than that of the pure element at given source temperature. In Fig.~\ref{fig_vap_pressure_all} the vapor pressure curves for a number of group III elements (a), group IV elements (b) and other commonly used elements (c) are shown (solid lines). In addition, their corresponding main suboxides are also shown (dashed lines). The typical range of vapor pressures used in MBE (LPCVD) sources is marked by the blued (yellow) shaded area and helps estimating the related cell temperatures. 
While for Al, Ba, Ti, and Pb suboxide evaporation will not be relevant at any practical source temperature, it can be dominant for B, Ga, In, Si, Ge, Sn, La, and Sb in a sufficiently high oxygen background during oxide MBE or LPCVD. In addition, the vapor pressures of the (sub)oxides of Mo, Nb, Ru, Ta, V, and W shown in Ref.~\citenum{adkison2020} are orders of magnitude higher than those of the corresponding elements.

\section{Summary and Conclusions}

In summary, we have shown, that a metal effusion cell with a Sn or Ga charge
produces a notable suboxide (SnO or Ga$_2$O) flux over a large temperature range when
exposed to the oxygen background present during oxide film growth in MBE. 
This suboxide flux can exceed the metal flux at sufficiently low cell temperature, for example, the one used for Sn doping or comparably low Ga$_2$O$_3$ growth rates.
A kinetic rate-equation-based model for the suboxide and metal flux was developed that takes into account the vapor pressure curves of metal and suboxide as well as the background oxygen pressure and metal-cell temperature. The model is shown to quantitatively describe our experimental data as well as published data on Sn-doping concentrations and Ga$_2$O$_3$ growth rates \cite{bierwagen2013a,han2018a,ahmadi2017,white2009a,simion2019,martinez2018a, kracht2017a,PCEickhoff2020,vogt2018a,vogt2017a,mazzolini2020, oshima2018,oshima2017,PCOshima2020,cheng2018a}.

For sufficiently low cell temperatures, it predicts the suboxide flux to increase with increasing cell temperature proportional to the suboxide vapor pressure curve before saturating in an oxygen-transport-limited plateau at higher cell temperatures. This saturated suboxide flux increases linearly with the background oxygen pressure. At sufficiently high cell temperature, the metal flux exceeds the suboxide flux and follows the metal vapor pressure curve. 
The resulting total (combined suboxide and metal) flux, that is relevant for growth, follows an $\textsf{S}$-shape as it moves from the suboxide-desorption limited regime, to the oxygen-transport-limited plateau, and further to the metal-desorption limited regime. Extrapolating to higher oxygen partial pressures, used in LPCVD of Ga$_2$O$_3$  and In$_2$O$_3$ , the model predicts the suboxide flux from the metallic source to be orders of magnitude higher than the metal flux under the growth conditions published in Refs.~\citenum{zhang2020,rafique2016,feng2019a,karim2018,zhang2019}.
As shown previously by Kalarickal $et$ al. for a Si source \cite{kalarickal2019a}, suboxide evaporation is not limited to Sn and Ga. Based on vapor pressure data of elements and their suboxides we predict suboxide evaporation to be potentially dominant for B, Ga, In, Sb, La, Ge, Si, Sn, Mo, Nb, Ru, Ta, V, and W, as their suboxide vapor pressure exceeds that of the element at given temperature. In contrast, suboxide evaporation is predicted to be irrelevant for Ba, Al, Ti, and Pb.

In the practical oxide-MBE use, the suboxide flux from elemental sources may be unexpected as it is absent during flux measurements (e.g by ion gauges) that are typically performed in the absence of an oxygen background. 
For metals with particularly low vapor pressure (e.g., Mo, Nb, Ru, Ta, V, and W), the suboxide evaporation can be an enabler for the MBE growth of their related (complex/multicomponent) oxides by avoiding the extreme source temperatures required to evaporate the metal. These multifunctional oxides (e.g., MoO$_3$, NbO$_2$, LiNbO$_3$, SrRuO$_3$, Ba$_2$RuO$_4$, Ta$_2$O$_5$, LiTaO$_3$, VO$_2$, V$_2$O$_5$, BiVO$_4$, SrVO$_3$, and WO$_3$) are highly relevant for (future) oxide electronics.\cite{Engel-Herbert2013, schlom2015,Lorenz2016,Coll2019}
At the same time, care needs to be taken about the oxidation and suboxide evaporation of Mo, Ta, or W-containing heated parts or filaments of effusion cells, substrate heaters, and substrate holders used in the oxide MBE chamber.

Suboxide evaporation can further be relevant at standby temperatures of the sources at which the elemental evaporation is negligible, and 
source oxidation can lead to a memory effect with non-stationary suboxide and metal fluxes. As a more stable and predictable alternative to elemental sources in oxide MBE, suboxide sources using oxide source charges or mixed elemental and oxide charges are recommended \cite{hoffmann2020a,adkison2020}. Such sources have already been used for Sn-doping \cite{sasaki2012}, BaSnO$_3$ growth \cite{raghavan2016}, or SnO$_2$\cite{hoffmann2020a} and Ga$_2$O$_3$ growth \cite{vogt2021}, respectively. 

\section*{Acknowledgments}
The authors thank Hongping Zhao, Patrick Vogt, Takayoshi Oshima, and Martin Eickhoff for sharing technical details of their growth with us. We further thank Steffen Behnke for the technical support
and maintenance work on the system, Christian Roethlein for his support
on the QMS system, and Thomas Auzelle for critically reading the manuscript.
This work was performed in the framework of GraFOx, a
Leibniz-ScienceCampus partially funded by the Leibniz
association. G.H. gratefully acknowledges financial support by the Leibniz-Gemeinschaft under
Grant No. K74/2017.

\section*{Data availability}
The data that support the findings of this study are available from the corresponding authors upon reasonable request.

\bibliography{qms_bib}
\end{document}